\documentclass[sigconf,nonacm]{acmart}

\usepackage{soul}
\usepackage{fontawesome5}
\usepackage{algorithm2e}
\usepackage{hyperref}
\usepackage{balance}
\usepackage{makecell}
\usepackage{siunitx}

\newcommand{\cmt}[1]{\ignorespaces}

\SetKwInput{KwInput}{Input}
\SetKwInput{KwOutput}{Output}
\SetKwInput{KwName}{Name}

\usepackage{mathtools}

\usepackage{multirow}

\AtBeginDocument{%
  }

\begin{document}

% \title[]{``Observe It, Draw It'': Design and Evaluation of an Observational Drawing System that Promotes Children's Connectedness to Nature}
% \title[Origami Interface]{Origami Interface: Rapid Fabrication of Electric, Tangible Input Interfaces with Haptic Feedback Using 4D Printed Origami}
% \title[Origami Interface]{Origami Interface: Rapid Fabrication of Haptic and Electric Input Interfaces Using 4D Printed Origami}
\title[TopoStyle]{TopoStyle: Supporting Iterative Design with Generative AI for 2.5D Topology Optimization}


\author{Shuyue Feng}
\orcid{0000-0002-6720-5356}
\affiliation{%
  \institution{The University of Tokyo}
  \city{Tokyo}
  \country{Japan}
}
\email{shuyuefeng@akg.t.u-tokyo.ac.jp}

\author{Cedric Caremel}
\orcid{0000-0002-3547-7285}
\affiliation{%
  \institution{The University of Tokyo}
  \city{Tokyo}
  \country{Japan}
}
\email{cedric@akg.t.u-tokyo.ac.jp}

\author{Yoshihiro Kawahara}
\authornote{Corresponding author.}
\orcid{0000-0002-3547-7285}
\affiliation{%
  \institution{The University of Tokyo}
  \city{Tokyo}
  \country{Japan}
}
\email{kawahara@akg.t.u-tokyo.ac.jp}

\renewcommand{\shortauthors}{Feng et al.}

\begin{abstract}
Topology optimization(TO) is widely used in engineering because of its ability to save material and optimize structural performance. Although prior work has explored 2D human-centered design tool for TO, the results are often limited in variety and offer weak customizability. Meanwhile, due to the high computational and time costs of TO, researchers have attempted to address these issues using generative AI; however, such methods often provide limited interactivity. In addition, topology optimization in many cases needs to balance structural performance and aesthetic qualities through iterative design, a perspective that has rarely been emphasized in traditional TO.
We present TopoStyle, an iterative design tool for 2.5D topology optimization using a 2D diffusion model. We explore two interaction methods. The first exports 3D parts to a graphical interface for hand-drawn interaction. The second enables direct interaction within 3D modeling software using points. Our tool also supports the use of masks to apply topology optimization to specific regions, allowing users to address customized design needs. We compare and evaluate both performance and interaction methods, and investigate how TopoStyle can balance performance and aesthetics while improving design efficiency through customization and iterative design. Finally, we demonstrate the application scenarios of TopoStyle through several design cases.
\end{abstract}

% <concept_id>10003120.10003121.10003125</concept_id>
% <concept_desc>Human-centered computing~Interaction devices</concept_desc>

% \ccsdesc[500]{Human-centered computing~Interaction devices}


 \begin{teaserfigure}
   \includegraphics[width=\textwidth]{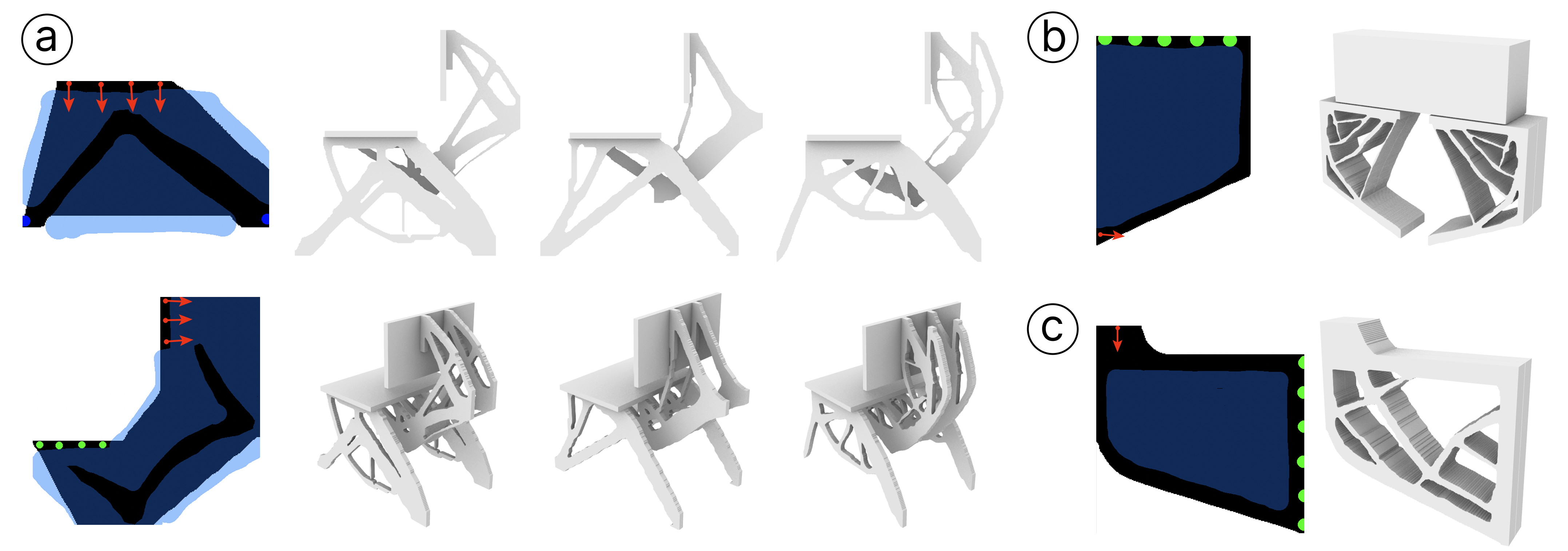}
   \caption{Applications created with TopoStyle: (a) a chair, with topology optimization applied to two different regions; (b) a robot gripper; and (c) a wall hook.}
   \Description{In Figure 1.}
  \label{Figure1}
 \end{teaserfigure}

\maketitle

\section{Introduction}

To achieve material efficiency and optimal structural performance in structural design, topology optimization (TO) has been widely adopted as a foundational technique~\cite{bendsoe2003topology, bendsoe1988generating, sigmund2013topology}. However, the hours-long convergence time of a single run in conventional topology optimization, along with its high computational cost, can reduce designers’ efficiency during early-stage exploration and limit the number of iterations they can perform~\cite{aage2017giga, yano2021globally}. In addition, due to the nature of the algorithm, topology optimization often produces only a narrow range of results~\cite{andreassen2011efficient}. Although Forte has introduced a sketch-based design interface, the customizability of the optimization outcomes remains limited~\cite{chen2018forte}. These issues constrain designers’ ability to satisfy structural requirements while also addressing design aesthetics. Yet in many scenarios, aesthetic qualities play a role that is as important as, or even more important than, performance and cost~\cite{loos2022towards}. 

On the other hand, some studies have begun to leverage generative AI to directly generate topology optimization results~\cite{giannone2023aligning,nobari2024nito, huang2022problem}. These data-driven approaches have been shown to produce TO structures with relatively high accuracy and much greater speed. However, from a practical usability perspective, they still suffer from clear limitations. First, many existing methods assume the entire image region, typically a square canvas, as the material domain to be optimized ~\cite{maze2023diffusion}. While this setting simplifies model training, it is difficult to adapt to real design tasks involving parts with specific geometric shapes. Second, the interaction mechanisms of existing systems remain unintuitive. Physical constraints such as loads and boundary conditions often need to be specified through parameter configuration or text-based embeddings, which increases interaction complexity and raises the barrier to use ~\cite{shneiderman1983direct}. Sketch2Topo has already explored the use of hand-drawn sketches to edit a 2D generative-AI-based topology optimization tool
~\cite{feng2026sketch2topo}. Although 2D topology optimization is commonly used in prior work and is sufficient for many benchmark and early-stage design scenarios, previewing results in a 3D view and reasoning about relationships among components remain important in practical design workflows.

To bridge these gaps, we present TopoStyle, an interactive iterative design tool for 2.5D topology optimization based on generative AI (diffusion models). It supports arbitrary shape inputs and can generate diverse topology optimization results under the same physical constraints while still satisfying structural performance requirements. We provide two interaction methods: one allows users to draw physical constraints directly on a 2D canvas using a brush, while the other enables users to control physical conditions by inserting points within 3D modeling software. In addition, we incorporate the mask and inpainting capabilities of diffusion models into 3D model editing~\cite{choi2021ilvr,saharia2022palette}, allowing users to select the regions where topology optimization should take place using geometry, thereby supporting a higher degree of customization.

We describe the usage workflow and implementation details of the tool. We also conducted evaluations including minimum compliance and volume fraction. We also conducted a comparative evaluation of the two TopoStyle workflows to demonstrate their improvements in efficiency and reduced interaction complexity. Furthermore, we explore applications to show that TopoStyle enables more efficient iterative topology optimization design through a simple interaction logic, while better supporting the balance between performance and aesthetics.

The main contributions of this paper are as follows:

\begin{itemize}
  \item An interactive tool, TopoStyle, that enables topology optimization with generative AI in 3D modeling software, using hand-drawn input to specify physical constraints and control the optimization region;
  \item Through quantitatively analyzing minimum compliance and volume fraction, we demonstrate the effectiveness of TopoStyle in terms of structural performance, and we further conduct a comparative study using KLM to show that it can improve efficiency and reduce interaction complexity;
  \item Three applications show that TopoStyle supports more efficient iterative topology optimization design with a simple interaction logic, while being more conducive to balancing performance and aesthetic qualities.
\end{itemize}

\section{Related Work}

\subsection{Topology Optimization}

Topology optimization (TO) is a classical method for obtaining high-performance structures by optimizing material distribution under given loads, boundary conditions, and material constraints~\cite{bendsoe2003topology,wu2015system}. Classical algorithms such as the Solid Isotropic Material with Penalization (SIMP) method~\cite{bendsoe1989optimal} and the Level Set Method~\cite{osher2001level} rely on iterative finite element analysis (FEA) to continuously update the physical field~\cite{turner1956stiffness}. Although these methods can produce structures with excellent mechanical performance, they are often accompanied by high computational complexity and long optimization times~\cite{aage2017giga}. To improve efficiency, researchers have begun introducing generative AI methods, ranging from early topology prediction based on Convolutional Neural Networks (CNN)~\cite{yu2019deep, sosnovik2019neural, abueidda2020topology}, to Generative Adversarial Networks (GAN)~\cite{nie2021topologygan}, and more recently to diffusion-model-based approaches that achieve higher-quality and more stable results ~\cite{giannone2023aligning, zhang2025research, nobari2025optimize}. TopoDiff was among the first to demonstrate that topology optimization tools trained with diffusion models significantly outperform GAN-based models in terms of minimum compliance, and that their performance shows real promise for practical engineering applications ~\cite{maze2023diffusion}. However, existing methods remain limited in interactivity: users can only input abstract physical parameters, such as loads and volume fractions, without understanding the generation process itself. This creates a black-box experience for users and makes it difficult to incorporate specific geometric constraints or perform localized modifications. Forte therefore developed a user-driven design tool based on traditional finite element analysis, which similarly uses sketch input to generate topology optimization results~\cite{chen2018forte}. Although pioneering, Forte relies on a deterministic solver and can produce only a single optimal output. By contrast, Sketch2Topo introduces sketch input into diffusion-model-based topology optimization, transforming physical constraints that previously had to be specified through parameters into inputs that can be expressed through hand drawing. Moreover, owing to the stochastic nature of diffusion models, it is able to generate more diverse results\cite{feng2026sketch2topo}. However, their interaction remains limited to an isolated 2D canvas that is later extruded into 2.5D, effectively stripping away the complex spatial context inherent in real 3D design tasks, such as relationships with surrounding parts ~\cite{sutton2007spatial}. In this work, we address this limitation by using diffusion models and embedding the generation process directly in situ within 3D modeling software for 2.5D generation, thereby turning it into a context-aware, divergent spatial dialogue.

\subsection{Control Mechanisms in Diffusion Models}

In engineering design, the value of diffusion models lies not only in the visual plausibility of their outputs, but also in whether they can support fine-grained control over structural form and local regions. However, generation methods that rely solely on text prompts are often unable to accurately express the requirements of engineering tasks regarding geometric boundaries, spatial relationships, and functional regions~\cite{liu2022design,chong2025prompting, avrahami2022blended, avrahami2023blended}. To improve controllability, researchers have proposed a range of conditional editing methods. For example, approaches such as ControlNet and T2I-Adapter introduce additional conditioning branches to inject spatial priors, such as sketches, edge maps, and depth maps, into the diffusion process, thereby improving the model’s responsiveness to structural signals~\cite{zhang2023adding,mou2024t2i}. At the same time, inpainting allows users to regenerate only specified masked regions, making localized modification possible~\cite{choi2021ilvr,saharia2022palette}; inversion methods such as DDIM inversion further support preserving the overall structure and original features of unedited regions during editing, thereby improving the consistency and stability of local adjustments~\cite{mokady2023null}. Nevertheless, these methods have largely focused on visual content control rather than functional structural design under performance constraints. In particular, within topology optimization, there remains limited exploration of how user input can be translated into effective control over material distribution and local structural intent. Therefore, we combine hand-drawn input, localized mask-based editing, and diffusion-based generation to enable users to participate more directly in the topology optimization process, thereby improving the customizability and interactivity of generative design.

\subsection{Structural Optimization and Aesthetic Qualities}

Structural optimization generally includes size optimization~\cite{haftka2012elements}, shape optimization~\cite{sokolowski1992introduction}, and topology optimization~\cite{bendsoe1989optimal}, among other approaches. Recently, researchers have discussed its aesthetic dimension: the branching, perforated, and skeletal forms produced by structural optimization are often regarded as embodying a distinctive performance-driven aestheticc~\cite{zhang2023machine, beghini2014connecting, oval2024similarity}. Some studies have also explicitly investigated how subjective preferences and aesthetic principles can be incorporated into topology optimization~\cite{li2023interactive, li2025interactive}. Such discussions are not limited to research; a series of products showcased through Autodesk Generative Design have also used topology optimization as a design reference~\cite{autodesk_gm_generative_design_2026}. Razer’s new Cobra Mini mouse likewise adopts a perforated design to reduce weight while adding visual appeal~\cite{razer_viper_mini_signature_2026}. These examples suggest that users often care simultaneously about performance, manufacturability, and visual style, rather than pursuing lightweighting alone. Despite this, traditional topology optimization still relies on relatively long solving and validation cycles, making it less suitable for high-frequency iteration and rapid form exploration driven by aesthetic preferences. Motivated by this gap, our work embeds diffusion models into 3D modeling software in order to support faster and more interactive exploration of performance–aesthetics trade-offs while retaining the benefits of structural optimization.

\section{Method}

\begin{figure}
\centering
\includegraphics[width=\linewidth]{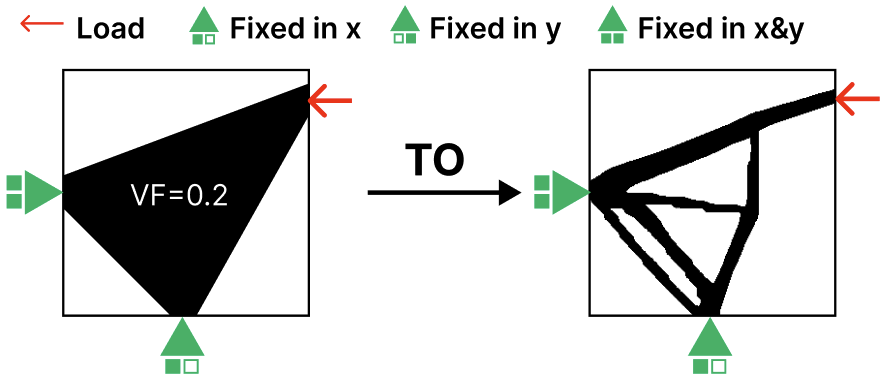}
\caption{Left figure shows the original shape of the material and the various input constraints. Right figure shows the result generated by topology optimization under these constraints.}
\label{Figure2}
\end{figure}

At the beginning, we first provide a brief introduction to how topology optimization works and to several commonly used variables. As shown in the \autoref{Figure2}, users first need to define basic parameters such as load locations, support conditions, and volume fraction. Based on these inputs, the system then generates a structural result that satisfies the required mechanical performance. The right side of the figure shows the optimized result under these constraints. By removing material that contributes less to the structure, topology optimization can produce structural forms that balance lightweighting and mechanical performance.

\subsection{Interaction Methods}

\begin{figure*}
\centering
\includegraphics[width=0.9\linewidth]{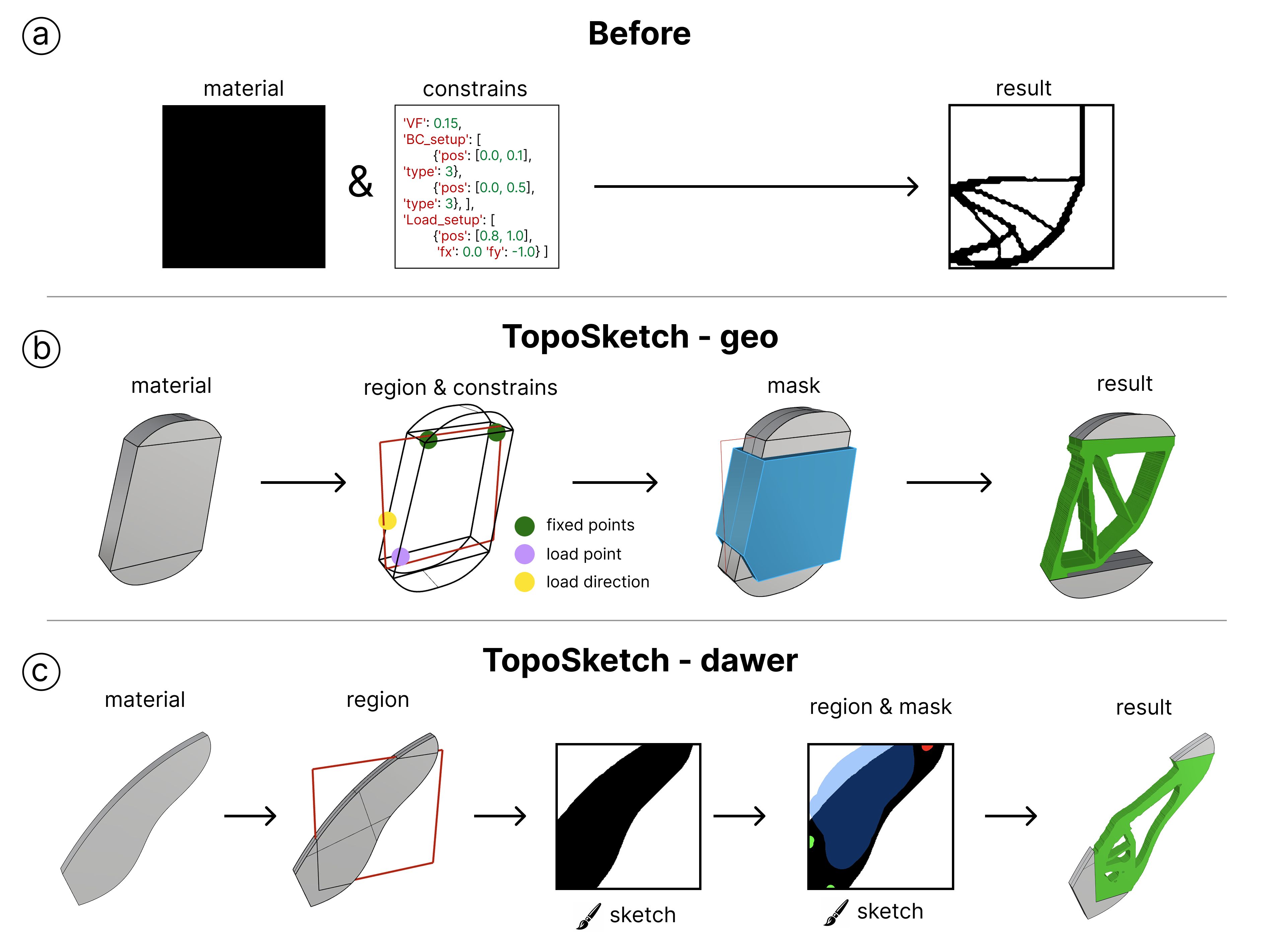}
\caption{Three different interaction methods for topology optimization using generative AI}
\label{Figure3}
\end{figure*}

In this section, we discuss prior work that has primarily focused on technical validation, and clarify how TopoStyle differs in terms of interaction design and what our work contributes.

Previous generative-AI-based topology optimization research has mainly focused on demonstrating feasibility, while giving limited attention to usability. To simplify model training, many variables were reduced or idealized for discussion ~\cite{maze2023diffusion}. For example, the optimized object was often simplified to a square block; in practice, however, target objects can have a wide variety of geometric forms. In addition, in conventional workflows, the input of physical constraints is entirely separated from the geometric object, which is not intuitive for users, as shown in \autoref{Figure3}a.

TopoStyle therefore introduces an image-to-image generation approach, allowing the object to be optimized to take arbitrary shapes. It also enables physical constraints to be specified by directly drawing on the part itself. In addition, it introduces a masking function that allows users to apply topology optimization only to selected regions.

Furthermore, TopoStyle aims to integrate 2.5D topology optimization into 3D modeling software, enabling users in multi-part scenarios to better perceive relationships among components and thereby improving the interaction experience. Users can create the objects to be optimized directly within the modeling software. For the subsequent interaction, we provide two approaches. The first, DRAWER (\autoref{Figure3}c), follows the sketch-based interaction paradigm of Sketch2Topo, but the base geometric shape in the canvas is converted from the 3D geometry created in the modeling environment, so users no longer need to hand-draw the base shape. Users only need to draw the physical constraints and masks in the GUI to perform iterative design. In addition, the generated results are directly reflected back into the 3D modeling software, allowing users to more intuitively inspect the overall design situation. The second, GEO (\autoref{Figure3}b), offers a more direct approach by using the modeling software itself to specify the physical conditions. We use points and vectors to define fixed supports and loading points. Other numerical parameters, such as volume fraction, are still entered as text. We also transplant the masking function directly into the modeling software: instead of the original mask representation, we use a geometric volume so that topology optimization occurs only within the region occupied by that geometry, thereby supporting a higher degree of customization.

In TopoStyle, we provide these two workflows not merely as competing alternatives, but as a fluid bimodal interaction framework designed to support different cognitive stages of the design process. The DRAWER method serves as a low-threshold interface for divergent exploration, allowing users to quickly sketch their intent without worrying about precise dimensions. Once a promising structural style has been identified, users can seamlessly transition to the GEO method for convergent refinement. Here, native 3D operations enable the precise specification of the exact physical constraints required for final engineering validation. This approach allows designers to move freely between rapid aesthetic brainstorming and precise parametric refinement within a single, unified context.

\subsection{Design Tool}

We used Rhino as the 3D modeling software and Grasshopper Python Script to transfer data and call local code. In the following, we describe the workflows and usage of the two methods we provide separately.

\begin{figure}
\centering
\includegraphics[width=\linewidth]{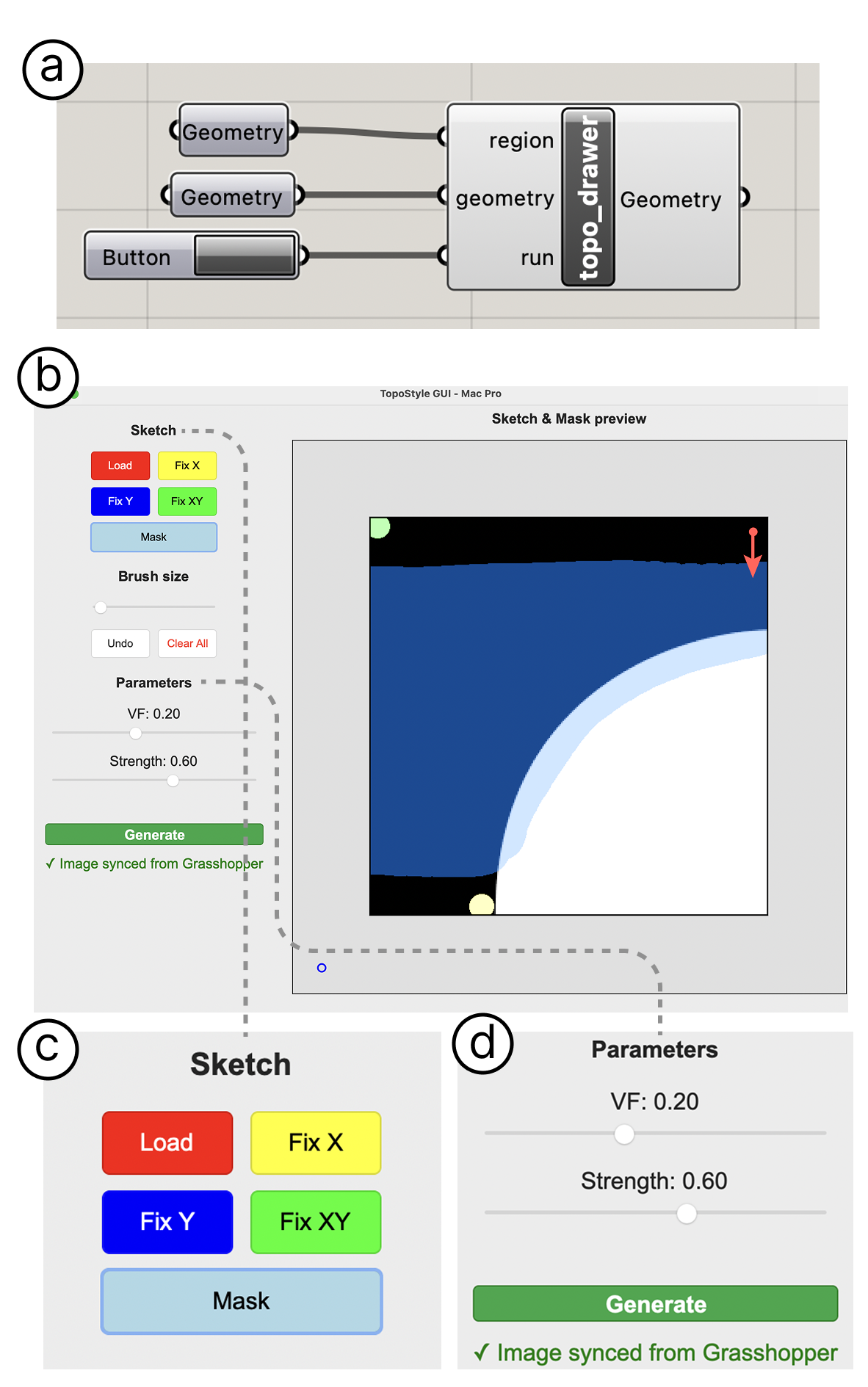}
\caption{User interface of TopoStyle-DRAWER. (a) The Grasshopper component used to connect Rhino and the drawing canvas. (b) The hand-drawing canvas interface. (c) and (d) Detailed views of interface functions.}
\label{Figure4}
\end{figure}

For the hand-drawn input method (DRAWER method), its core component is a Grasshopper component named topo\_drawer (\autoref{Figure4}a). Users first create the geometry to be optimized in the 3D modeling software (\autoref{Figure3}d material), and then create a square to define the region for topology optimization(\autoref{Figure3}c region). These inputs are then connected to the DRAWER method component and executed. Next, the user only needs to open the GUI, and the geometry in the 3D modeling software is automatically converted into a planar graphic and displayed on the canvas(\autoref{Figure4}b). Different brush colors represent different constraint conditions. We improved the input method for loads: instead of specifying the loading direction through text, users now use a red arrow to indicate both the loading point and loading direction. In addition, yellow indicates a fixed point constrained only in the x-direction, blue indicates a fixed point constrained only in the y-direction, green indicates a fixed point constrained in both x and y, and cyan represents the mask(\autoref{Figure4}c). The volume fraction and denoising strength are entered as text(\autoref{Figure4}d). After completing the setup, users click generate, and the result is directly reflected in the 3D modeling software.

\begin{figure}
\centering
\includegraphics[width=\linewidth]{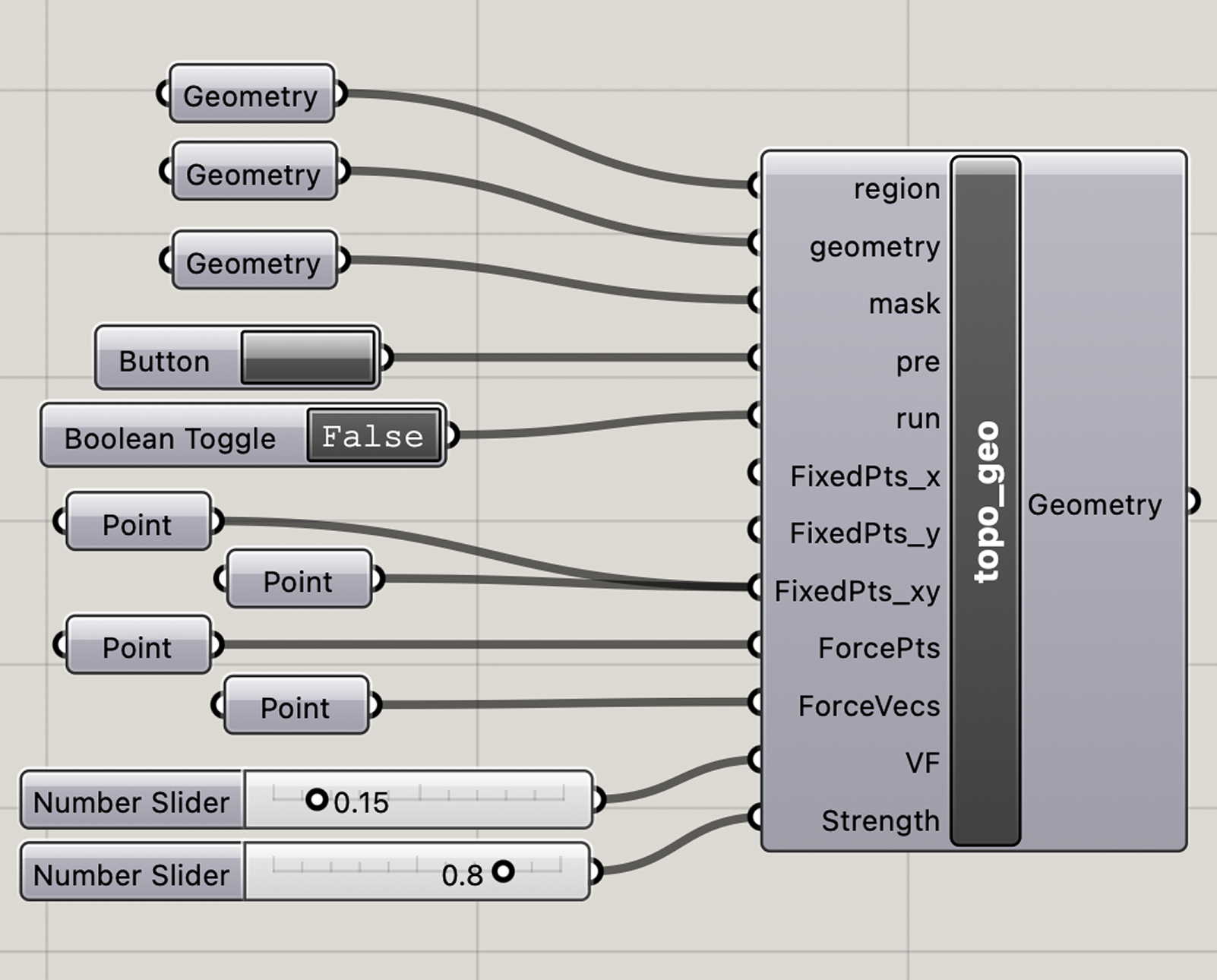}
\caption{Grasshopper components of TopoStyle-GEO.}
\label{Figure5}
\end{figure}

For the geometry-based input method, its core component is a Grasshopper component named topo\_geo (\autoref{Figure5}). Similar to the hand-drawn method, users first create the required geometry and a square (to define the generation region) and import them into Grasshopper. Next, users create a series of points in Rhino to represent fixed points, loading points, and loading directions (\autoref{Figure3}c material). The data of these points are also connected to the component. Since the volume fraction and strength parameters are numerical by nature, they are directly entered into the component as text. In addition, users can create extra geometry as a mask input (\autoref{Figure3}c mask), whose function is to designate a selected region so that topology optimization only occurs within that region. Because all inputs are geometric information, users can make precise modifications and perform iterative design more easily.

\subsection{Technical Details}

First, regarding the diffusion model itself, TopoStyle does not require retraining the base model; instead, it only modifies the diffusion logic of the underlying model. This makes it easy to replace the current model with a stronger one in the future. We use the open-source TopoDiff as our base model~\cite{maze2023diffusion}. To simplify training, many parameters are fixed. For example, the filter radius is set to 2.0, Poisson’s ratio to 0.3, and the penalty factor to 3.0.

To enable the model to support specific geometric shapes, we incorporate an image-to-image strategy into the base model. The basic idea is to first add a certain amount of noise to the input image and then denoise it again, without requiring model fine-tuning~\cite{saharia2022palette}. For the hand-drawn input workflow, we use the same approach as Sketch2Topo: computer vision techniques based on OpenCV are used to recognize colored regions as corresponding coordinates~\cite{opencv_library}, which are then converted into a standard format and passed to the diffusion model. For the geometric input workflow, the coordinate information of the input points is extracted directly and normalized. All of these functions are implemented in Python and communicate through the Python Script component in Grasshopper within Rhino.

\section{Evaluation}

\subsection{Structural Performance Evaluation}

\begin{figure}
\centering
\includegraphics[width=\linewidth]{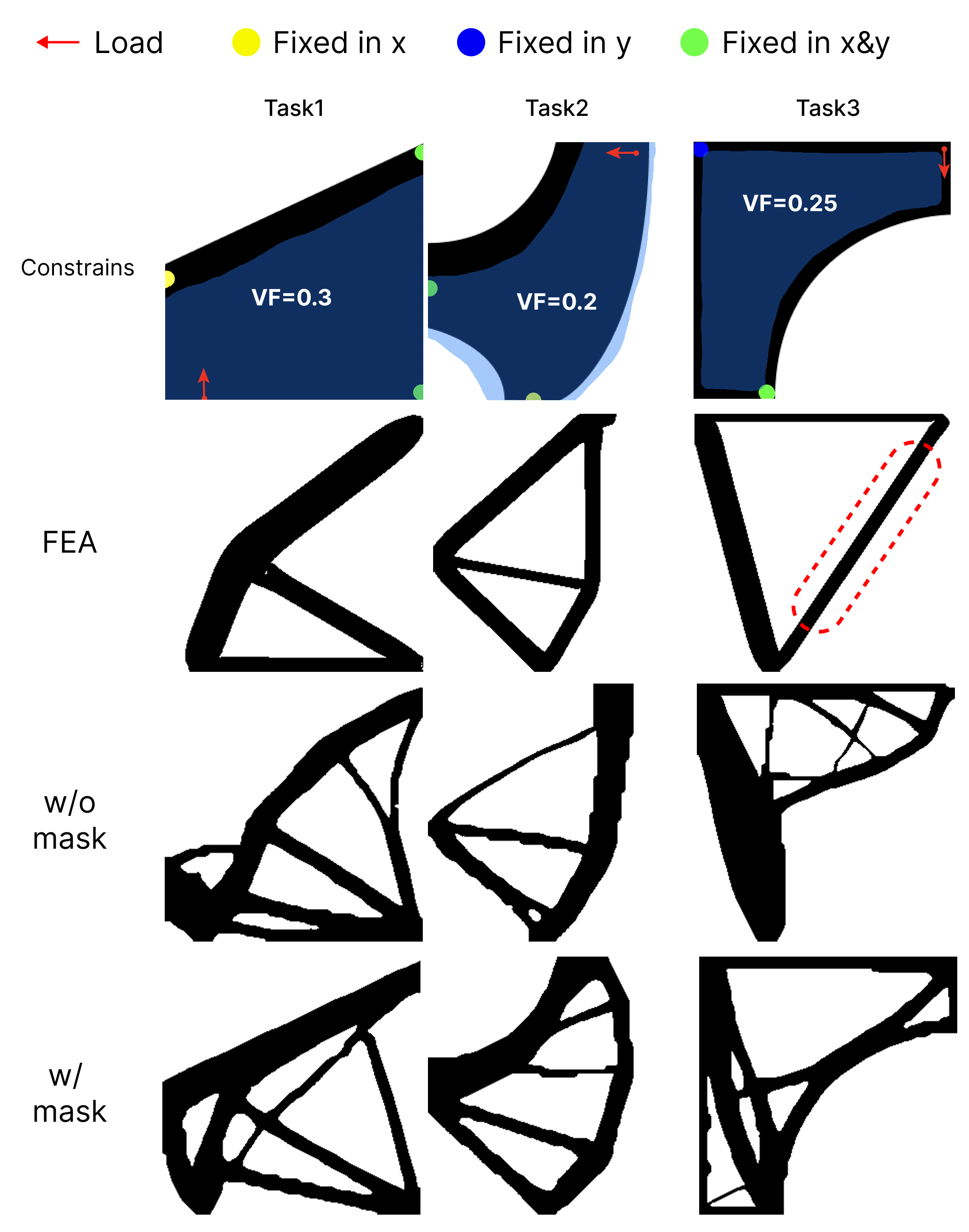}
\caption{Physical boundary conditions of the three tasks in the performance evaluation, along with previews of the generated results.}
\label{Figure6}
\end{figure}

To verify that TopoStyle is not only visually similar to topology optimization results but can also satisfy the performance requirements of topology optimization, we conducted a quantitative evaluation of two key properties: \textit{minimum compliance} and \textit{volume fraction}. We defined three different tasks and generated results using three methods: the traditional finite-element-analysis-based approach (ground truth), TopoStyle without mask, and TopoStyle with mask\autoref{Figure6}. For TopoStyle, we generated 20 samples under each condition and reported the average results. The conditions for each task are shown in the figure. In addition, all other parameters were kept consistent: the filter radius was set to $r_{\mathrm{min}} = 2.0$, Poisson's ratio to $\nu = 0.3$, the penalty factor to $p = 3.0$, and the Young's modulus to $E_0 = 1$ for the solid phase and $E_{\mathrm{min}} = 10^{-9}$ for the void phase. The results are shown in ~\autoref{tab1}.

\begin{table*}[t]
\centering
\caption{Comparison of minimum compliance and volume fraction across three tasks.}
\label{tab1}
\begin{tabular}{llccc}
\hline
 &  & Task1 & Task2 & Task3 \\
\hline
\multirow{3}{*}{Minimum Compliance} 
& FEA      & 21.15        & 60.33         & 144.96        \\
& w/o Mask & 26.03 $\pm$ 4.17  & 59.81 $\pm$ 13.16 & 237.86 $\pm$ 51.56 \\
& w/ Mask  & 23.15 $\pm$ 1.83  & 52.18 $\pm$ 8.30  & 186.54 $\pm$ 19.15 \\
\hline
\multirow{3}{*}{Volume Fraction (\%)} 
& FEA      & 30           & 20            & 20            \\
& w/o Mask & 34.44 $\pm$ 1.08 & 25.32 $\pm$ 0.92 & 30.98 $\pm$ 0.47 \\
& w/ Mask  & 37.05 $\pm$ 0.87 & 28.24 $\pm$ 0.75 & 36.01 $\pm$ 1.09 \\
\hline
\end{tabular}
\end{table*}

Overall, the generated results for Task 1 and Task 2 are very close to the ground truth in terms of minimum compliance and remain within an acceptable range. However, their volume fractions are noticeably higher than the ground truth. We can see that when a mask is used, the volume fraction deviates even more from the ground truth, because after introducing the mask, the preserved material regions contribute little to the mechanical performance. As a result, if TopoStyle attempts to maintain the target volume fraction during generation, it must sacrifice structurally effective regions, which in turn increases the minimum compliance. Conversely, if it prioritizes maintaining a low minimum compliance, the volume fraction must increase. Therefore, by adjusting the guidance strength of the base model, we prioritize minimum compliance, since this directly determines the structural performance of the material. If we want results closer to the ground truth in terms of volume fraction, we can reduce the volume fraction parameter during generation, but this will also degrade structural performance accordingly. We also note that the minimum compliance in Task 2 is even lower than the ground truth, mainly because its volume fraction is larger. Put simply, more material is used, so the structural performance naturally improves. In contrast, Task 3 shows a clear increase in minimum compliance and greater variation in volume fraction. Through analysis, we found that the selected shape constrained the model from generating toward the optimal solution, because part of the optimal structure lies outside the prescribed geometry (\autoref{Figure6} Task 3, FEA). This means that no matter how well the model performs, it cannot reach the true optimum under this shape constraint. This issue is therefore unrelated to the performance of the model or the tool itself.

In addition, we observed that in Task 1, the fixed point on the far left is not connected in the ground truth either. This is because, under these boundary conditions, that point has no tendency to move and therefore does not require any supporting material. This is a relatively special case. When examining all results generated by TopoStyle, we found that some results, like the ground truth, contain no material at that point, while others do. We consider this to be a positive behavior, because the user’s intention in selecting that point as a fixed point may be to encourage a particular structural shape rather than to pursue the purely optimal topology optimization solution.

\subsection{System Performance Evaluation}

\begin{table}[t]
\centering
\caption{KLM operator definitions and estimated execution times.}
\label{tab2}
\begin{tabular}{lll}
\hline
Operator & Description & Time (s) \\
\hline
$\mathbf{K}$    & Keystroke or button press & 0.20 \\
$\mathbf{P}$    & Moving the mouse cursor to the target & 1.10 \\
$\mathbf{H}$    & Homing between different devices & 0.40 \\
$\mathbf{D}$    & Drawing & 6.70 \\
$\mathbf{M}$    & Mental preparation & 1.35 \\
$\mathbf{R}_{t1}$ & Response time of the Grasshopper & 1.00 \\
$\mathbf{R}_{t2}$ & Model inference response time & 10.00 \\
\hline
\end{tabular}
\end{table}

To investigate the differences between the two workflows we proposed, we used the Keystroke-Level Model (KLM) to quantitatively evaluate their interaction time~\cite{card1980keystroke}. In brief, KLM estimates how long an expert user takes to complete a routine task and accounts for what they are doing during that time. The model operates at the level of primitive actions, such as pressing a key or moving a pointing device. The duration of each action is listed in \autoref{tab2}. Among them, $\mathbf{K}$, $\mathbf{P}$, $\mathbf{H}$, and $\mathbf{M}$ follow the standard KLM definitions, while $\mathbf{D}$, $\mathbf{R}_{t1}$, and $\mathbf{R}_{t2}$ were obtained from our own tests.

\begin{figure}
\centering
\includegraphics[width=\linewidth]{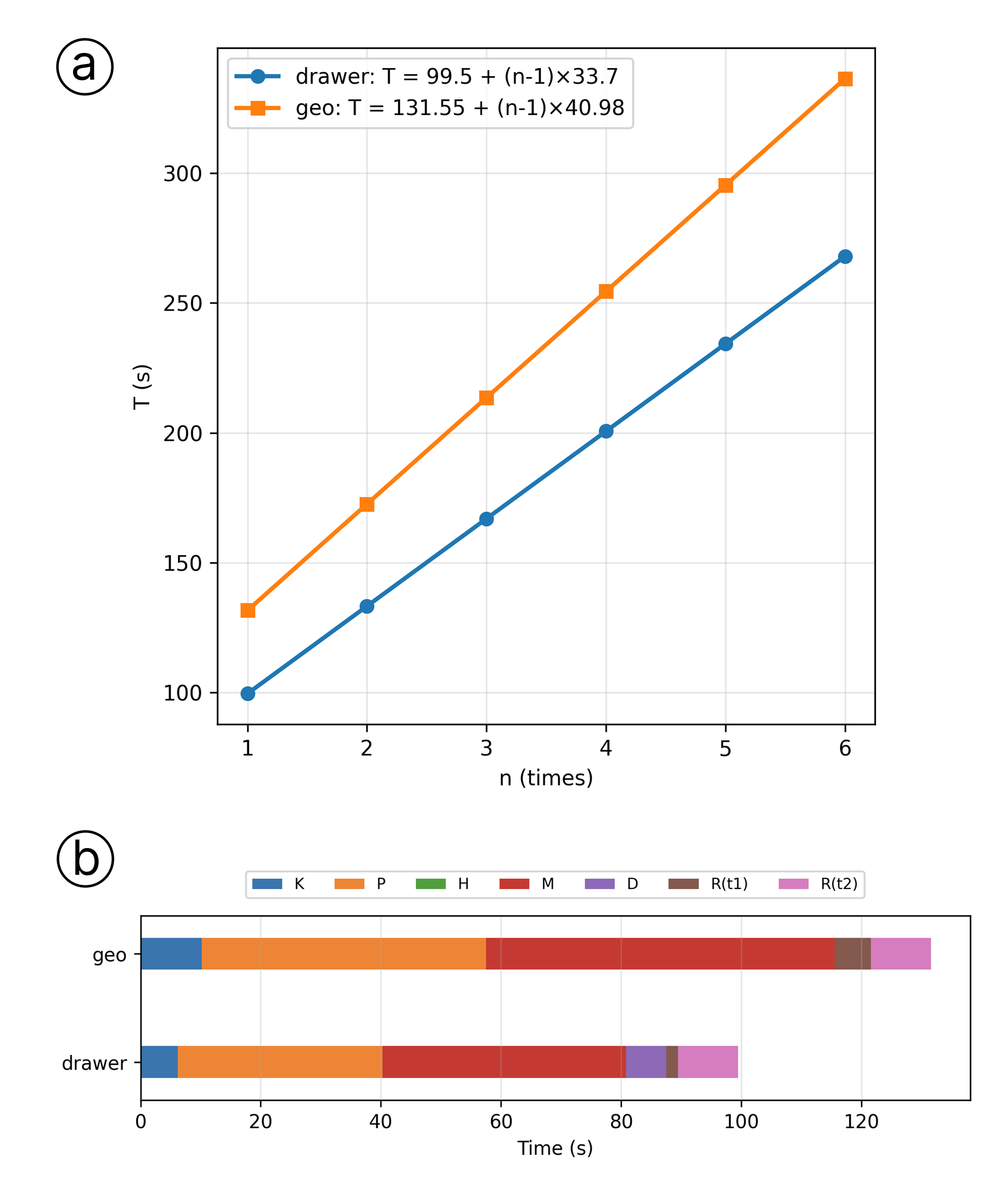}
\caption{(a) Time required by the two interaction methods across multiple iterations. (b) Proportion of time spent on different actions during a single generation task for the two methods.}
\label{Figure7}
\end{figure}

We conducted a comparative experiment by having the two workflows perform exactly the same task. Specifically, the task consisted of one topology optimization run followed by one iteration. The optimization target, load points, fixed points, and mask were kept identical across the two conditions. By decomposing the task into primitive steps, we derived formulas relating total interaction time to the number of iterations for the two workflows, as shown in \autoref{Figure7}a. The detailed step-by-step breakdown is provided in \autoref{sssec:drawer_workflow_task}.

As shown in \autoref{Figure7}a, the DRAWER method workflow requires less total time than GEO method regardless of the number of iterations. The interaction style of GEO method is closer to traditional 3D topology optimization tools, such as those in Fusion 360 and SolidWorks. This suggests that the sketch-based DRAWER method interface is more efficient than performing all editing directly in the 3D environment, and that it also maintains a lower incremental cost in subsequent iterations. In other words, the advantage of DRAWER method is not limited to the initial setup, but extends to iterative design exploration.

\autoref{Figure7}b further shows the time distribution of different operation types for a single design cycle ($n=1$). Crucially, the DRAWER method involves substantially less mental preparation time ($\mathbf{M}$) than the GEO method. This dramatic reduction in cognitive friction is a direct empirical validation of our underlying system architecture. By automating the non-trivial bidirectional translation between 3D parametric contexts and 2D pixel-based AI, TopoStyle successfully eliminates the tedious operational overhead (exporting, aligning, context-switching) that typically interrupts a designer’s flow. Instead of constantly calculating how to translate aesthetic intent into rigorous parametric operations, users can focus purely on ``intent externalization'' and ``what-if'' spatial reasoning ~\cite{cox1999representation, goldschmidt1991dialectics}. Nevertheless, GEO method also has its own value. Because it relies on geometric input, it enables more precise specification of optimization conditions and is therefore necessary in application scenarios where dimensional accuracy is critical. In such cases, a practical strategy may be to first use DRAWER method for rapid multi-round exploration and then use GEO method in the final stage for precise refinement. More broadly, the advantage of DRAWER method lies in reducing the cost of externalizing design intent. Instead of specifying optimization conditions entirely through parameterized 3D operations, users can communicate local structural intent through sketch-based interaction.

\section{Application}
To demonstrate the advantages of TopoStyle in iterative design and customization, we present three applications. The chair case demonstrates its ability to support multi-round iterative design. The robot gripper case demonstrates customized design under different functional requirements. The wall hook case demonstrates its ability to support customization through shape control. For clarity in presenting the physical conditions, we use TopoStyle-DRAWER throughout this section.

\subsection{Chair}

\begin{figure*}
\centering
\includegraphics[width=\linewidth]{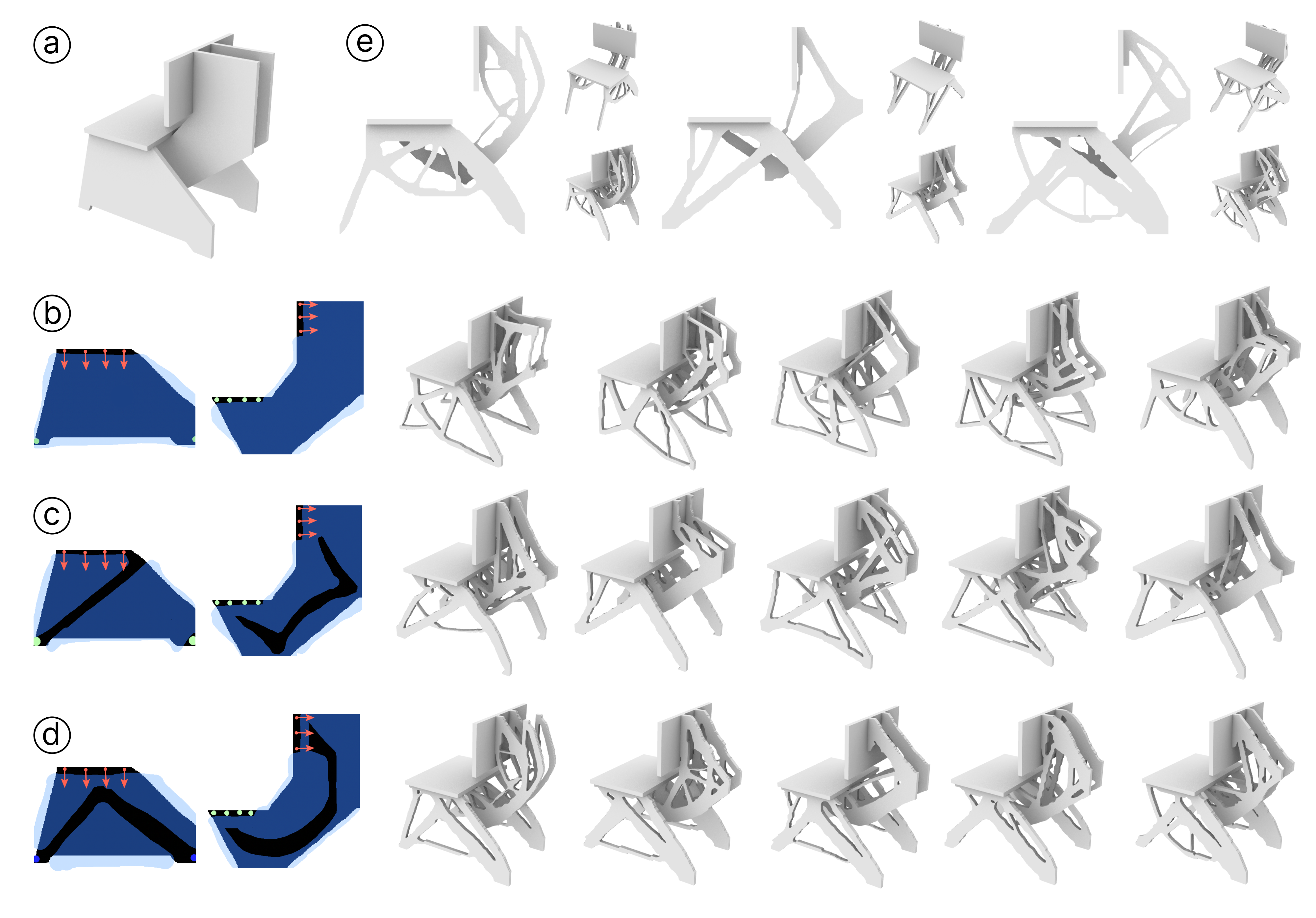}
\caption{Iterative design of a chair using TopoStyle: (a) the base 3D model of the chair; (b) the constraints used for the first topology optimization; (c) and (d) topology optimization results after modifying the mask; and (e) three chairs with different design styles selected from the generated results.}
\label{Figure8}
\end{figure*}

We created a base 3D model, as shown in \autoref{Figure8}a. We separately applied topology optimization to two of its components: the connector between the backrest and the seat, and the chair leg. We first performed unbiased full-region topology optimization to extract design inspiration, as shown in \autoref{Figure8}b. In total, we generated 20 results and selected 5 of them for presentation. Based on the resulting chair forms as inspiration, we then adjusted the mask shapes for iterative design, as shown in \autoref{Figure8}c and d. Because the two components were generated independently, they can be freely combined according to user preference. Finally, we selected three chair designs, as shown in \autoref{Figure8}e.

\subsection{Robot Gripper}

\begin{figure}
\centering
\includegraphics[width=\linewidth]{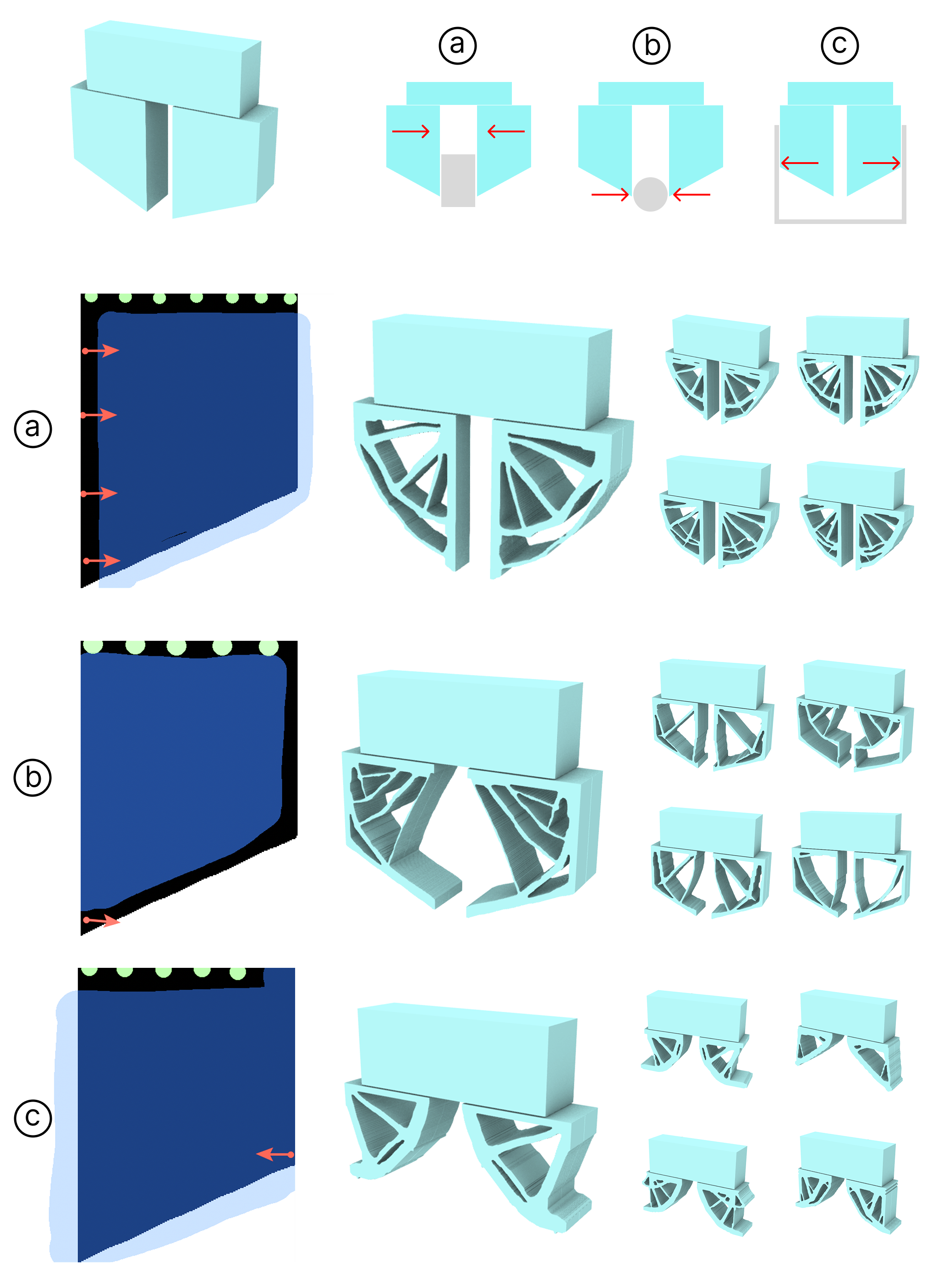}
\caption{Designing structures with different functions for a robot gripper using TopoStyle: (a) for grasping large objects; (b) for grasping and securely holding objects with low friction; and (c) for outward expansion tasks.}
\label{Figure9}
\end{figure}

The base 3D model of the robot gripper and the three different grasping tasks are shown in \autoref{Figure9}. The first task is to pick up a large object. Therefore, we applied topology optimization by setting the left side of the robot gripper as the load-bearing surface (taking the right half of the robot gripper as an example), and the result is shown in \autoref{Figure9}a. The second task is to pick up an object with a smaller contact area, which implies lower friction. Because the friction is too small, simply gripping the object may cause it to slip. Therefore, we require the gripper not only to hold the object, but also to contain an internal cavity that can store it. The result is shown in \autoref{Figure9}b. The third task is outward expansion. In this case, the robot gripper is inserted into the object and expands outward to grip the inner wall so as to lift the object. Therefore, we placed the load point on the right side for topology optimization, and the result is shown in \autoref{Figure9}c.

\subsection{Wall Hook}

\begin{figure}
\centering
\includegraphics[width=\linewidth]{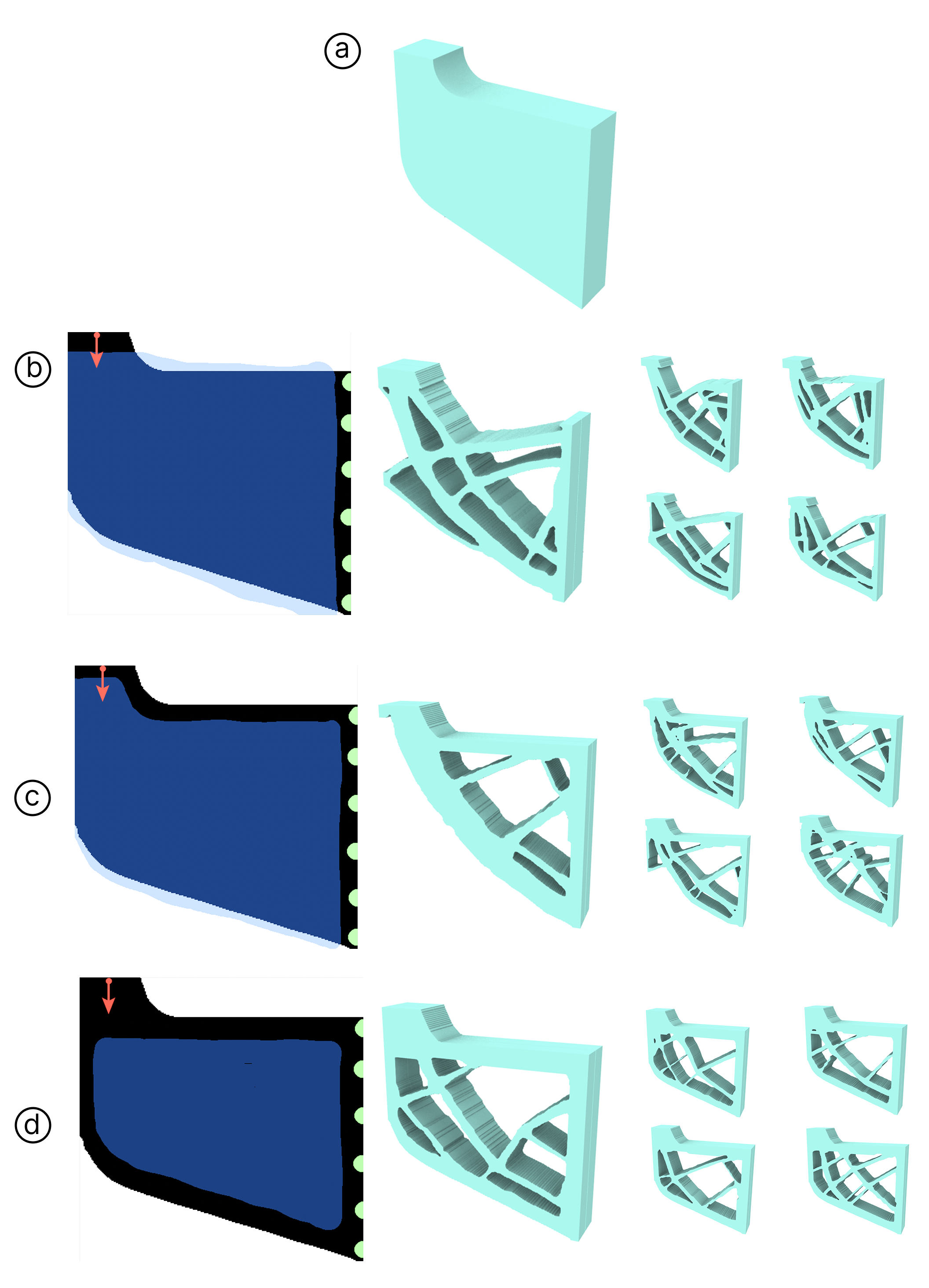}
\caption{Shape design of a wall hook using TopoStyle: (a) the base 3D model of the wall hook; (b) topology optimization over the entire region; (c) preserving the top shape; and (e) preserving the entire outer contour.}
\label{Figure10}
\end{figure}

The base 3D model of the wall hook is shown in \autoref{Figure10}a. The constraint input shown in \autoref{Figure10}b is similar to that of traditional topology optimization: it only specifies the load region and the fixed region, while the remaining parts are generated entirely by the algorithm. In contrast, TopoStyle allows users to preserve additional regions beyond these basic constraints in order to satisfy shape preferences. For example, users may choose to preserve the upper half while performing topology optimization, as shown in \autoref{Figure10}c, or preserve the entire outer contour while performing topology optimization, as shown in \autoref{Figure10}d.
\section{Discussion, Limitations, and Future Work}

\subsection{From One-Shot Solving to Iterative Design Exploration}

The value of TopoStyle lies not only in shortening the setup time for a single topology optimization task, but also in reducing the marginal interaction cost of continuous iteration, thereby changing the role of topology optimization in the design process. Traditional topology optimization tools are typically closer to one-shot engineering solvers: users must explicitly define geometry, constraints, and loads before waiting for the system to return a result. In contrast, TopoStyle reorganizes topology optimization as an interactive form-finding medium that can be embedded into the 3D modeling process, allowing users to continuously propose, adjust, and compare local structural alternatives during design. In particular, the lower mental preparation time and lower incremental iteration cost observed in the DRAWER method workflow suggest that its advantage lies not only in shortening the initial setup time, but also in supporting repeated and continuous design exploration. In this sense, TopoStyle is not merely a plugin for accelerating optimization, but an interaction framework that brings generative structural optimization earlier into the design exploration stage. However, this conclusion is currently based mainly on KLM analysis. While KLM helps compare the operational cost and marginal interaction cost of different workflows, it cannot substitute for long-term studies in real design contexts. Future work should therefore examine more directly how the system affects designers’ exploration breadth, iteration frequency, and decision-making processes.

\subsection{Sketch-Based Interaction as Low-Threshold Intent Expression}

Another key contribution of TopoStyle lies in how it reorganizes the expression of design intent. The lower mental preparation time observed in the DRAWER method workflow suggests that the advantage of a sketch-based interface is not merely that it is “more intuitive,” but that it reduces the cost of externalizing design intent into system operations. Rather than relying entirely on parameterized 3D operations to express local modification intent, users can participate in structural generation through the more direct representational form of sketching. This finding further suggests that, in topology optimization, performance and aesthetics are not simply in opposition, but can be negotiated and rebalanced through repeated iterative interaction. In contrast, although the GEO method workflow is less efficient, it provides greater precision through geometric input and therefore remains irreplaceable in scenarios where dimensional accuracy is critical. At the same time, sketch input also has clear limitations. While it lowers the barrier to expression, it may introduce ambiguity in local structural intent. For example, different users may use similar strokes to express different thicknesses, support logics, or boundary preferences. This suggests that sketch interaction is better understood as a low-threshold means of expressing intent, rather than as a complete substitute for explicit constraint definition. Future work could further explore richer hybrid workflows, such as combining sketch input with precise geometric input, or introducing mixed-reality interfaces that allow users to move more flexibly between low-threshold expression and high-precision control.

\subsection{Technical Scope}

Although TopoStyle demonstrates the potential of combining diffusion models with 2.5D topology optimization, its current technical pipeline still has clear limitations. First, the system does not perform generation directly in a native 3D volumetric space (e.g., dense voxel grids), but instead supports 3D workflows through a 2.5D (planar extrusion in-situ) diffusion-based editing mechanism. This was a deliberate HCI and systems design choice: while true 3D volumetric AI generation represents an ultimate technical frontier, it currently suffers from prohibitive computational costs that break interactive, real-time human-in-the-loop design. Because a vast majority of practical engineering components—such as brackets, ribs, and robotic end-effectors—fundamentally rely on planar topologies, our 2.5D in-situ pipeline provides a highly effective, pragmatic balance between computational speed (sub-minute generation) and spatial context awareness. As a result, the current system remains limited in handling complex multi-directionally coupled constraints and intricate internal 3D cavities. Second, the current performance evaluation and design cases mainly focus on objects of moderate complexity under relatively standardized physical conditions, and do not yet cover more complex manufacturing constraints, dynamic loads, multi-material conditions, or real engineering deployment scenarios. Accordingly, the results of this paper are better understood as design support for early-stage structural exploration and interactive form-finding, rather than as a complete replacement for traditional engineering topology optimization workflows. Looking forward, future work could explore direct generation and editing in native 3D representations such as meshes or implicit representations, thereby reducing the information loss caused by 2D-to-3D mapping and improving the system’s ability to represent complex spatial structures and local constraints. Further incorporating manufacturing constraints, fabrication feedback, and physical testing into a closed loop would also help advance this type of tool from interactive support in the conceptual stage toward a more complete generative manufacturing workflow. More broadly, the greater potential of TopoStyle lies not in becoming a single optimization tool, but in serving as a new interaction interface that more tightly connects generative AI, structural optimization, and design practice.
\section{Conclusion}
This paper presents TopoStyle, a generative-AI-based iterative design tool for 2.5D topology optimization. Unlike traditional topology optimization tools, which are mainly used as one-shot engineering solvers, TopoStyle reframes topology optimization as an iterative design tool that supports more direct structural generation and local modification in 3D modeling environments. We explored two workflows: a sketch-based DRAWER method workflow and a point-based GEO method workflow, both of which support localized optimization and customized design exploration. Our evaluation examined both structural performance and interaction efficiency. The results show that TopoStyle can generate topology optimization outcomes with comparable mechanical performance under given physical conditions. At the same time, the KLM analysis indicates that the DRAWER method workflow requires less total operation time and lower incremental iteration cost than the GEO method workflow, suggesting that sketch-based interaction is better suited for rapid and repeated design exploration. Together with the design cases, these findings show that TopoStyle can support not only performance-oriented optimization, but also local refinement and aesthetic exploration. tyleOverall, TopoStyle is not merely a plugin that applies diffusion models to topology optimization. Rather, it rethinks generative structural optimization as an interactive design medium in 3D modeling. We hope this work can inspire future research at the intersection of generative AI, structural optimization, and design practice.

\balance
\bibliographystyle{ACM-Reference-Format}
\bibliography{references}

\newpage
\appendix

\section{Tasks Used for KLM Analysis}
Task Design: Optimized topology on an object with a mask, two loading points, and four fixed points using two method

\label{sssec:drawer_workflow_task}

\begin{description}
    \item[TopoStyle-DRAWER]
    \item[A.]
    \textit{Open Grasshopper and the plugin.} \\$(MPK)\times 4$ \\
    Move the mouse $\rightarrow$ Click Grasshopper $\rightarrow$ Move the mouse $\rightarrow$ Click to open the file $\rightarrow$ Move the mouse $\rightarrow$ Click the software $\rightarrow$ Move the mouse $\rightarrow$ Click Open.

    \item[B.]
    \textit{Draw the topology optimization region.}\\ $MPK + MPK + MP$ \\
    Move the mouse $\rightarrow$ Click the button $\rightarrow$ Move the mouse $\rightarrow$ Click $\rightarrow$ Draw the region.

    \item[C.]
    \textit{Input geometry into Grasshopper.}\\ $(MPK + MPK + MPKKR_{t1})\times 2$ \\
    Move the mouse $\rightarrow$ Click Select Object $\rightarrow$ Move the mouse $\rightarrow$ Click $\rightarrow$ Move the mouse $\rightarrow$ Click $\rightarrow$ Click $\rightarrow$ System response $R_{t1}$ $\rightarrow$ Repeat once, then perform the same sequence again.

    \item[D.]
    \textit{Draw the sketch.}\\ $(MPK)\times 12 + MPKD + MPKP$ \\
    Move the mouse $\rightarrow$ Click Run $\rightarrow$ Move the mouse $\rightarrow$ Click the load-point function $\rightarrow$ Move the mouse $\rightarrow$ Click and draw loading point 1 $\rightarrow$ Move the mouse $\rightarrow$ Click to specify the loading direction $\rightarrow$ Move the mouse $\rightarrow$ Click and draw loading point 2 $\rightarrow$ Move the mouse $\rightarrow$ Click to specify the loading direction $\rightarrow$ Move the mouse $\rightarrow$ Click the fixed-point $xy$ function $\rightarrow$ Move the mouse $\rightarrow$ Click and draw fixed point 1 $\rightarrow$ Move the mouse $\rightarrow$ Click and draw fixed point 2 $\rightarrow$ Move the mouse $\rightarrow$ Click and draw fixed point 3 $\rightarrow$ Move the mouse $\rightarrow$ Click and draw fixed point 4 $\rightarrow$ Move the mouse $\rightarrow$ Click the mask function $\rightarrow$ Move the mouse $\rightarrow$ Click $\rightarrow$ Move and draw the mask $\rightarrow$ Move the mouse $\rightarrow$ Click VF $\rightarrow$ Adjust by dragging.

    \item[E.]
    \textit{Run the program.}\\ $MPKR_{t2}$ \\
    Move the mouse $\rightarrow$ Click Generate $\rightarrow$ System response $R_{t2}$.

    \item[F.]
    \textit{Redraw in the $n$th iteration.}\\ $(MPK)\times 3 + MPKD + MPKP + MPKR_{t2}$ \\
    Clear the content $\rightarrow$ Move the mouse $\rightarrow$ Click Clear $\rightarrow$ Draw the sketch again $\rightarrow$ Move the mouse $\rightarrow$ Click the mask function $\rightarrow$ Move the mouse $\rightarrow$ Click $\rightarrow$ Move and draw the mask $\rightarrow$ Move the mouse $\rightarrow$ Click VF $\rightarrow$ Adjust by dragging $\rightarrow$ Run the program.

    \item[G.]
    \textit{Export the result.}\\ $(MPK)\times 2$ \\
    Move the mouse $\rightarrow$ Right-click the Grasshopper component $\rightarrow$ Move the mouse $\rightarrow$ Click Bake.
\end{description}

\label{sssec:geo_workflow_task}

\begin{description}
    \item[TopoStyle-GEO]
    \item[A.]
    \textit{Open Grasshopper and the plugin.} \\$(MPK)\times 4$ \\
    Move the mouse $\rightarrow$ Click Grasshopper $\rightarrow$ Move the mouse $\rightarrow$ Click to open the file $\rightarrow$ Move the mouse $\rightarrow$ Click the software $\rightarrow$ Move the mouse $\rightarrow$ Click Open.

    \item[B.]
    \textit{Draw the topology optimization region.}\\ $MPK + MPK + MP$ \\
    Move the mouse $\rightarrow$ Click the button $\rightarrow$ Move the mouse $\rightarrow$ Click $\rightarrow$ Draw the region.

    \item[C.]
    \textit{Draw the constraints.}\\ $(MPK)\times 9$ \\
    Move the mouse $\rightarrow$ Click the point function $\rightarrow$ Move the mouse $\rightarrow$ Click to place loading point 1 $\rightarrow$ Move the mouse $\rightarrow$ Click to place loading direction 1 $\rightarrow$ Move the mouse $\rightarrow$ Click to place loading point 2 $\rightarrow$ Move the mouse $\rightarrow$ Click to place loading direction 2 $\rightarrow$ Move the mouse $\rightarrow$ Click to place fixed point 1 $\rightarrow$ Move the mouse $\rightarrow$ Click to place fixed point 2 $\rightarrow$ Move the mouse $\rightarrow$ Click to place fixed point 3 $\rightarrow$ Move the mouse $\rightarrow$ Click to place fixed point 4.

    \item[D.]
    \textit{Draw the mask.}\\ $(MPK)\times 6$ \\
    Move the mouse $\rightarrow$ Click the rectangle drawing function $\rightarrow$ Move the mouse $\rightarrow$ Click to start creating the rectangle $\rightarrow$ Move the mouse $\rightarrow$ Click to finish the rectangle $\rightarrow$ Move the mouse $\rightarrow$ Click the extrude function $\rightarrow$ Move the mouse $\rightarrow$ Click to select the curve $\rightarrow$ Move the mouse $\rightarrow$ Click to generate.

    \item[E.]
    \textit{Input geometry into Grasshopper.}\\ $MPK + MPK + MPKKKKKR_{t1} + (MPK + MPK + MPKKR_{t1})\times 5$ \\
    Move the mouse $\rightarrow$ Right-click $\rightarrow$ Move the mouse $\rightarrow$ Click $\rightarrow$ Move the mouse $\rightarrow$ Click $\rightarrow$ Click $\rightarrow$ Click $\rightarrow$ Click $\rightarrow$ Click $\rightarrow$ System response $R_{t1}$ $\rightarrow$ Move the mouse $\rightarrow$ Right-click $\rightarrow$ Move the mouse $\rightarrow$ Click $\rightarrow$ Move the mouse $\rightarrow$ Click $\rightarrow$ Click $\rightarrow$ System response $R_{t1}$ $\rightarrow$ Repeat Steps 12--19 steps five times.

    \item[F.]
    \textit{Run the program.}\\ $MPKR_{t2}$ \\
    Move the mouse $\rightarrow$ Click Run $\rightarrow$ System response $R_{t2}$.

    \item[G.]
    \textit{Redraw in the $n$th iteration.}\\ $MPKHK + (MPK)\times 6 + (MPK + MPK + MPKKR_{t1}) + MPKR_{t2}$ \\
    Clear the content $\rightarrow$ Move the mouse $\rightarrow$ Click $\rightarrow$ Switch device $\rightarrow$ Click Delete $\rightarrow$ Draw the mask $\rightarrow$ Re-input the mask $\rightarrow$ Run the program.

    \item[H.]
    \textit{Export the result.}\\ $(MPK)\times 2$ \\
    Move the mouse $\rightarrow$ Right-click the Grasshopper component $\rightarrow$ Move the mouse $\rightarrow$ Click Bake.
\end{description}

\end{document}